# Performance of Hemielliptic Dielectric Lens Antennas with Optimal Edge Illumination

Artem V. Boriskin, Ronan Sauleau, and Alexander I. Nosich

***Abstract*— The role of edge illumination in the performance of compact-size dielectric lens antennas (DLAs) is studied in accurate manner using a highly efficient algorithm based on the combination of the Muller's boundary integral equations and the method of analytical regularization. The analysis accounts for the finite size of the lens and directive nature of the primary feed placed close to the center of the lens base. The problem is solved in a two-dimensional (2-D) formulation for both *E*- and *H*-polarizations. It is found that away from internal resonances that spoil the radiation characteristics of DLAs made of dense materials, the edge illumination has primary importance. The proper choice of this parameter helps maximize DLA directivity, and its optimal value depends on the lens material and feed polarization.**

***Index Terms*— Beam collimation, dielectric lens antenna, directivity, edge illumination, edge taper.**

## I. Introduction

Similarly to parabolic dish reflectors, elliptical dielectric lenses have the ability to collect rays propagating parallel to their axis of symmetry into their focus, e.g. [1, 2]. Reciprocally, in emitting mode, this shape is expected to provide a locally plane wave in the radiating aperture of the hemielliptic DLA. Note that while the whole reflector surface is involved into the beam forming, it is only the frontal part of elliptical lens surface that plays similar role; the rear part of lens, behind the central cross-section, has no importance (as suggested by the geometrical optics). A truly plane wave can emerge only for an infinite reflector or lens and omnidirectional source. For realistic feeds and antennas, however, the spillover and illumination losses are inherently present: the former are associated with the power that misses reflector or lens whereas the latter are due to a non-uniform illumination of the reflector or the lens front part, as schematically shown in Fig. 1. In reflector antenna theory it is usually stated that a -10 dB edge taper provides reasonable compromise for these losses and enables one to use the reflector in the most efficient way (see, e.g. [1], Fig. 4-4 and discussion thereafter).

In the case of DLAs, there is apparently no such analysis in open textbooks or research papers. Although there is a number of papers describing the performance of DLAs with variable lens extensions (e.g. see [3-5]), the role of edge illumination seems to have escaped a clear physical interpretation. Several obvious circumstances make a similar design advice for DLAs far from obvious. First, both the electrical size and the focal distance of elliptical DLAs are usually much smaller than that of reflectors [3], and thus the feed is never far away from the lens (as it is common for reflectors). Second, unlike a gently curved metal reflector, any dielectric lens is, in fact, an open dielectric resonator which is capable of supporting resonant modes. The quality factors of such modes depend on the lens parameters (shape, size, and permittivity) and can achieve rather high values for lenses made of dense materials such as silicon. If excited, internal resonances strongly affect the performance of DLAs [6, 7]. Finally, for DLAs, the focal distance and thus the favorable feed location depend on the lens material. This happens because, in geometrical optics approximation, the eccentricity of elliptical lens is determined by its material permittivity [2, 4]. These essential distinctions between reflector antennas and DLAs make the -10 dB optimal edge taper a questionable recommendation and call for additional study aimed at clarification of the role of edge illumination.

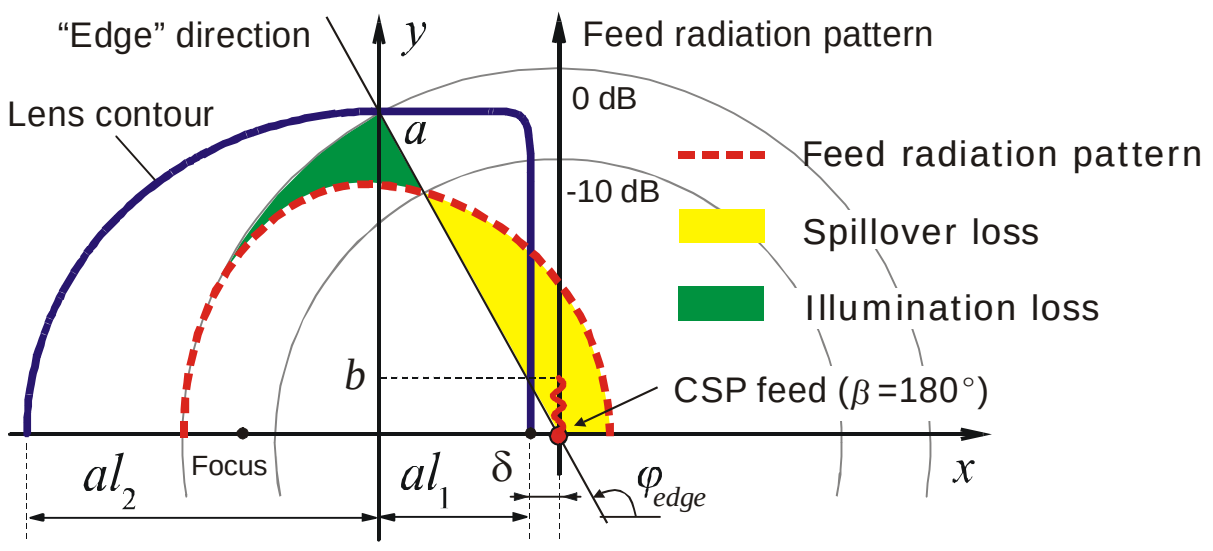

Fig. 1. Geometry and notations for the 2D model of a hemielliptic DLA (due to the symmetry only upper half is drawn). Curved line indicates the branch-cut appearing in the real space due to the CSP modelling the aperture feed. A schematic diagram for the CSP radiation pattern is given to interpret the losses associated with a non-uniform illumination of the lens front part.

To handle this problem, we study numerically the 2-D model of a compact-size hemielliptic DLA typically used in millimeter (mm) and sub-mm wave applications [3]. The lens is fed by an aperture source simulated using the complex source point (CSP) beam [8]. The analysis is performed using a highly-efficient algorithm based on the Muller boundary integral equations (MBIE) combined with the method of analytical regularization and Galerkin projection on the set of trigonometric polynomials [9]. It fully accounts for the resonance properties of the lens and guarantees uniqueness of the solution as well as fast convergence and controllable accuracy of the numerical algorithm. Details about this technique and its numerical implementation have been already given in [6,7] and therefore are not presented here. Comparing to [10] where performance of a hemielliptic silicon lens with varying extension is studied, this paper addresses a general problem of gaining the maximum efficiency of hemielliptic DLAs by adjusting the edge illumination level.

The definition of the edge illumination is given in Section II. Then in Section III we investigate the beam collimation properties of extended hemielliptic DLAs for various lens sizes and materials and feed apertures. Finally, the conclusions are drawn in Section IV.

## II. Mathematical Model

The lens is modelled by a homogeneous dielectric cylinder whose contour is combined from two curves, namely one half-ellipse whose eccentricity equals the inverse of the material refracting index ($e = 1/\varepsilon^{1/2}$, $l_2 = [\varepsilon / (\varepsilon - 1)]^{1/2}$), and one half-superellipse (rectangle with rounded corners), smoothly joint at the points $(x,y) = (0, \pm a)$, where $a$ is the minor semi-axis of the ellipse (Fig. 1). Hereafter, these points are referred as the "edge" of the lens aperture because the focusing ability of the lens is determined by its elliptical front part whereas the extension is used to place the feed at the right (focal) distance.

The feed is simulated by a complex source point (CSP) that is a current line located in complex space [8]. CSP is an attractive model

Manuscript received Nov. 11, 2008. Revised Jan. 22, 2009.

This work was supported in part by the joint projects of the National Academy of Sciences of Ukraine (NASU) with the Centre National de la Recherche Scientifique (CNRS), Ministère de l'Education Nationale, de l'Enseignement Supérieur et de la Recherche, and Ministère des Affaires Etrangères et Européennes, France. The first author was also supported by the Foundation Michel Métivier and by the NATO-RIG grant.

A. V. Boriskin and A. I. Nosich are with the Institute of Radiophysics and Electronics NASU, vul. Akad. Proskury 12, Kharkiv 61085, Ukraine (e-mail: a_boriskin@yahoo.com).

R. Sauleau is with the "Groupe Antennes et Hyperfréquences", Institut d'Electronique et de Télécommunications de Rennes, Université de Rennes 1, UMR CNRS 6164, 35042 Rennes cedex, France.

of aperture feeds because its field is a unidirectional beam whose waist is controlled by the value of the imaginary part of the CSP coordinate; it behaves like a Gaussian beam in the paraxial zone, whereas in the far zone the CSP field smoothly transforms into a cylindrical wave and thus (in contrast to a Gaussian beam) satisfies the Sommerfeld radiation condition at infinity [11]. For DLAs, the feed is usually fixed directly to the lens flat bottom. To account for this, the CSP is assumed radiating in a medium with the same permittivity as the lens, $\varepsilon$. Thus, its far-field asymptotic is given by

$$U(r,\varphi)_{r\to\infty} \sim \left(2/i\pi k_e r\right)\exp\left(ik_e r\right)\exp[k_e b\cos(\varphi-\beta)] \qquad (1)$$

where $k_e = 2\pi\sqrt{\varepsilon}/\lambda_0$ is the wavenumber in the medium, $b$ is the imaginary part of the CSP coordinate, and $\beta$ is the main beam direction.

For reflector antennas having large electrical size, the edge illumination is typically defined as the ratio in source power radiated in the edge direction and in the broadside [1]: $A = 20\log\left(U(\varphi_{edge})/U(\varphi_{bdside})\right)|_{r\to\infty}$, where $U$ is the field amplitude (in 2-D, this corresponds to $E_z$ and $H_z$ components for the *E*- and *H*-polarizations, respectively) and $\varphi_{bdside}$ is the broadside (forward) direction. Such a definition, based on the far-field radiation pattern of the feed, is convenient because it clearly explains the physical origins of the losses and simplifies engineering specifications for feeds. For hemielliptic lens fed by the CSP located and oriented as shown in Fig. 1, a closed-form expression for the edge illumination is defined by

$$A \approx -8.68\ k_e b\,(1+\cos\varphi_{edge}) \qquad \text{[in dB].} \qquad (2)$$

If the lens is cut through its rear focus then the normalized lens extension $l_1 = (\varepsilon-1)^{-1/2}$ and $\cos\varphi_{edge} = \varepsilon^{-1/2}$.

In spite of its prevalence for reflector antennas, the far-field definition of edge illumination is less applicable for hemielliptic DLAs whose typical size is only several wavelengths. Therefore it should be replaced by the one based on the near-fields, i.e. defined as the ratio of the incident field intensity at the "edge" of the lens aperture to that at the center of the aperture, i.e.

$$A = 20\log[U(0,a)/U(0,0)] \qquad \text{[in dB]} \qquad (3)$$

where $U(x,y)$ is the field amplitude at $(x,y)$.

The difference between these two definitions will be highlighted in Figs. 2, 5, and 6 (Section III) where two curves for the edge illumination defined via the far and near fields are indicated.

The radiation characteristic considered as a measure of collimation ability of the lens is the broadside directivity defined as $D = 2\pi\,|U(\varphi_{bdside})|^2/P_{tot}$, where $P_{tot} = \int_0^{2\pi}|U(\varphi)|^2 d\varphi$ is proportional to the radiated power integrated over all directions. Note that the directivity of the CSP radiating into infinite medium is $D_e = \exp(2\sqrt{\varepsilon}kb)/I_0(2\sqrt{\varepsilon}kb)$, where $I_0$ is the modified Bessel function. This function is represented by the curve marked with black circles in Figs. 2, 5, and 6 (Section III).

Finally it is important to notice that the same level of edge illumination can be provided by feeds with different radiation patterns. Therefore, the results given in Section III for the CSP feed, still require correction for other feeds depending on their radiation patterns.

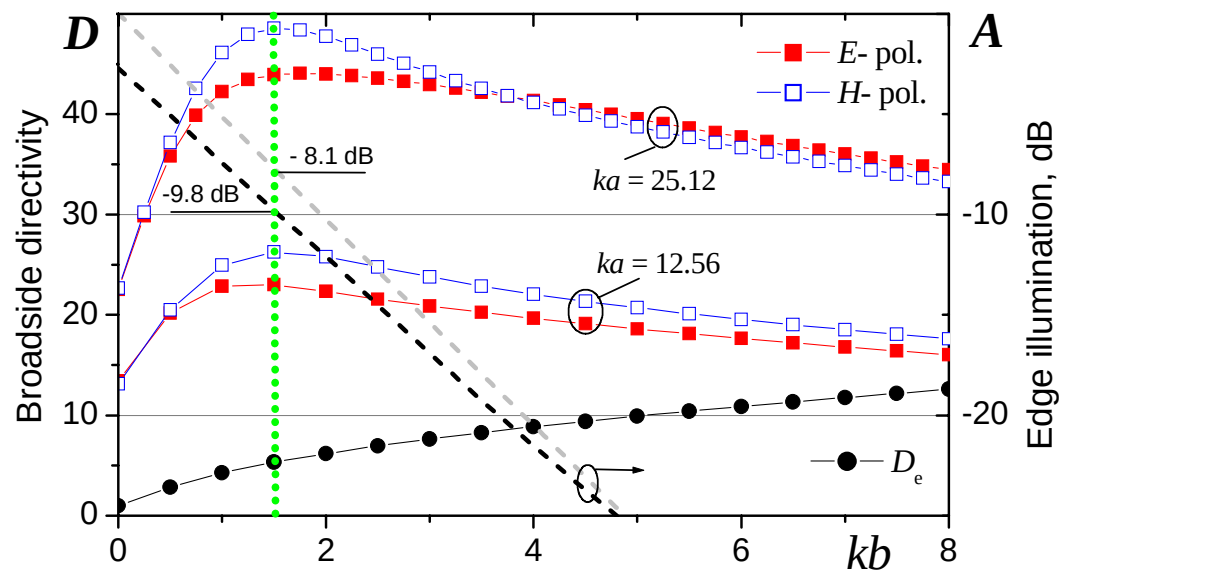

Fig. 2. Main beam directivity (left scale) and aperture edge illumination (right scale) of the cut-through-focus rexolite hemielliptic DLA ($l_1 = 0.8$, $l_2 = 1.286$) vs. CSP aperture width. The inclined dashed lines indicate the aperture edge illumination defined via the near- (black) and far-field (grey) formulation. The vertical dotted line indicates the maximum value of the directivity and is plotted to help estimate the optimal values of edge illumination.

## III. Numerical Results

The numerical results presented here include data for lenses of two different sizes (namely, *ka* = 12.56 and *ka* = 25.12; *k* is the free space wavenumber) and made in rexolite ($\varepsilon = 2.53$), isotropic quartz ($\varepsilon = 3.8$), and silicon ($\varepsilon = 11.7$). The lenses are illuminated by *E*- and *H*-polarized CSP feeds located symmetrically outside the lens at the distance $\delta = \lambda_0/20$ from its flat-bottom center. These parameters are typical for common mm and sub-mm applications and are beyond the range of reliable applicability of high-frequency methods [12].

### A. Rexolite lens

Fig. 2 represents the broadside directivity of rexolite hemielliptic DLA (left scale) and the corresponding edge illumination level (right scale) versus the parameter *kb* that controls the CSP virtual aperture. Here two pairs of curves, marked by filled and hollow squares, correspond to two lenses of different size and illuminated by the *E*- and *H*-polarized CSPs, respectively. The inclined dashed lines associated with the right axis show the level of edge illumination as given by Eqs. (2) and (3). As one can see, the maximum directivity for the cut-through-focus rexolite hemielliptic lenses of both sizes and in both polarizations is achieved under the same edge illumination, which is close to -10 dB (if defined via near field) or -8 dB (if defined via far field); the difference appears due to the reasons discussed above. Note, however, that the maximum is broad so any edge illumination between -8 and -12 dB will be acceptable.

More complete information about the collimation properties of rexolite DLAs can be extracted from the relief maps of the broadside directivity computed for lenses with variable extension size fed by a CSP with variable aperture (Fig. 3). Note that edge illumination depends on the both parameters, so the scale given in Fig. 2 corresponds only to the value of $l_1$ marked with vertical dashed lines. The non-monotonic behavior of directivity proves important role of internal reflections in the electromagnetic behavior of compact-size dielectric lenses. This highlights the fact that the accurate description of lens properties is only achievable with a full account for all wave phenomena even for low-index materials such as rexolite.

Fig. 3 reveals that a good choice of the lens extension itself does not guaranty the highest directivity. This value can be considerably improved by tuning the feed radiation and hence obtaining the optimal level of the lens edge illumination. The correspondence between the values of the *kb* parameter, the edge illumination for the cut-through-focus lens, and the directivity of the CSP in the medium can be extracted from Fig. 2, whereas the relief maps given in Fig. 3 can be used for finding the optimal relation between the feed and lens parameters needed to provide the highest directivity of DLAs.

The far-field radiation patterns of the three DLA configurations identified in Fig. 3(c) are presented in Fig. 4: B corresponds to the hemielliptic lens illuminated by an omnidirectional line current, whereas A and C are DLAs demonstrating the highest value of directivity at broadside when excited by CSP feeds providing the optimal edge illumination. Comparison of the curves reveals that a proper choice of the combination of the lens extension and the feed aperture enables one to reduce both backward radiation and side-lobes level.

### *B. Quartz and silicon lenses*

The variations of the broadside directivity $D$ versus the CSP aperture width, $kb$, for hemielliptic lenses in quartz (Fig. 5) and silicon (Fig. 6) reveal a behavior similar to the one presented above. The directivity grows until some optimal edge illumination is achieved and then almost monotonically goes down. If $kb$ is large, $D$ approaches a value which is below the directivity of CSP in the media: this can be explained by the backreflections of the feed radiation from the lens front and rear surfaces.

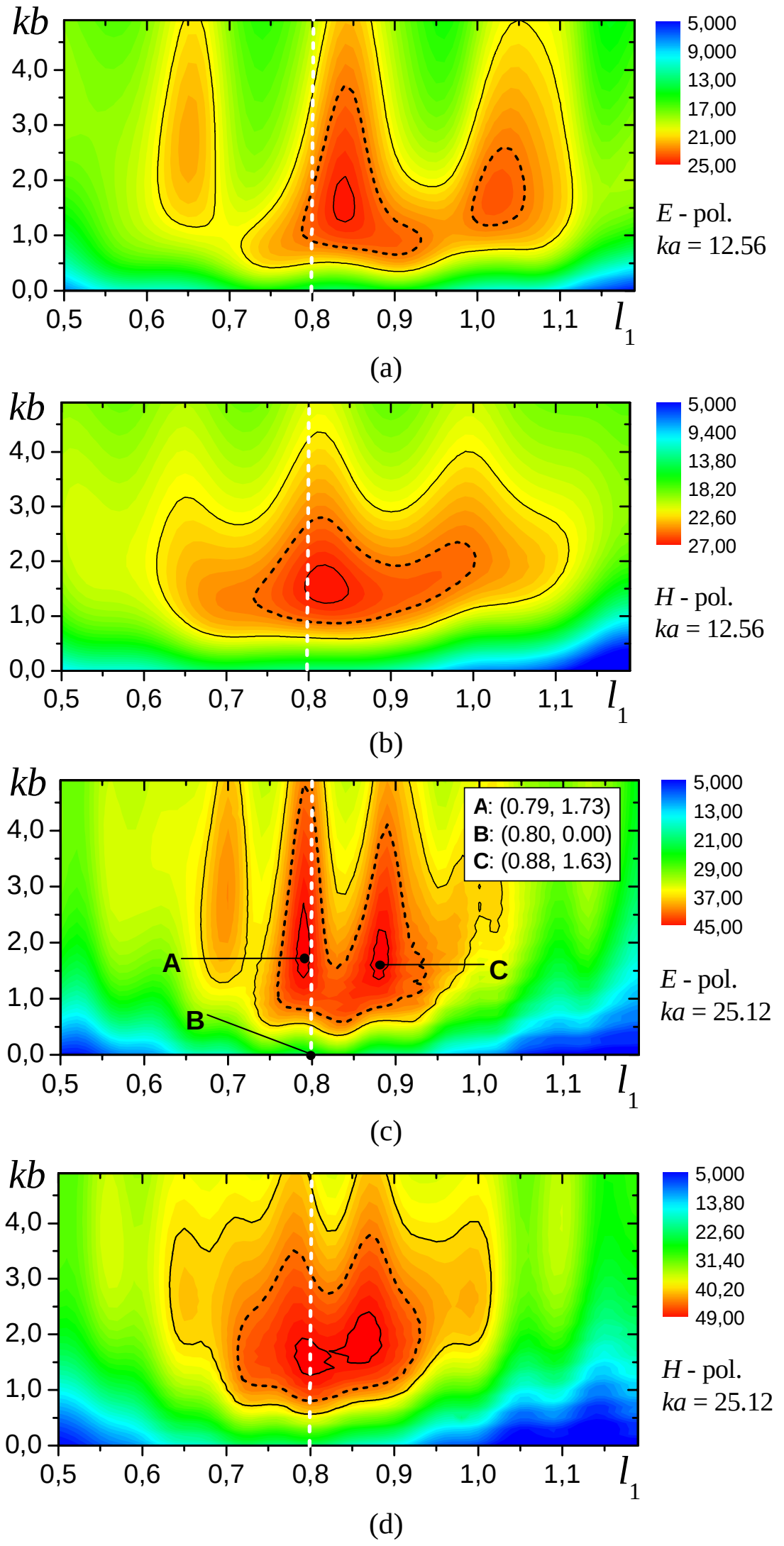

Fig. 3. Broadside directivity versus normalized lens extension ($l_1$) and CSP aperture width ($kb$). For clarity, only top 2%, 10%, and 20% value contours are shown. The white dashed lines indicate the length of the lens extension corresponding to the cut-through-focus design. The marks A, B, C in (c) correspond to the far-field radiation patterns given in Fig. 4.

Comparison of Figs. 5 and 6 with Fig. 2 shows that for lenses made of denser materials a greater difference appears between the values of the optimal edge illuminations for the $E$ and $H$-polarizations of the incident field. Depending on the lens size, it varies between -6.2 and -12.5 dB for quartz lenses, and between -10.5 and -23.5 dB for silicon ones, if defined via near fields. The far-field definition provides values higher by 3 and 5 dB for quartz and silicon lenses, respectively. The difference between polarizations that grows proportionally to material refraction index is explained by the growing role of internal reflections that are more pronounced in the $E$-case [11]. Note that the average (for different polarizations) value of the optimal near-field edge illumination for rexolite and quartz lenses is at the same level of around -10 dB, whereas the far-field value grows from -8 to -7 dB. This happens because for elliptical lenses the focal distance depends on permittivity of the lens material. Due to this, the angle $\varphi_{edge}$ decreases proportionally to the lens permittivity (see Fig. 1). As a result, the impact of the illumination loss grows for denser materials whereas the effect of spillover loss decreases. In such a way, a compromise between both types of losses is found at different values of edge illumination.

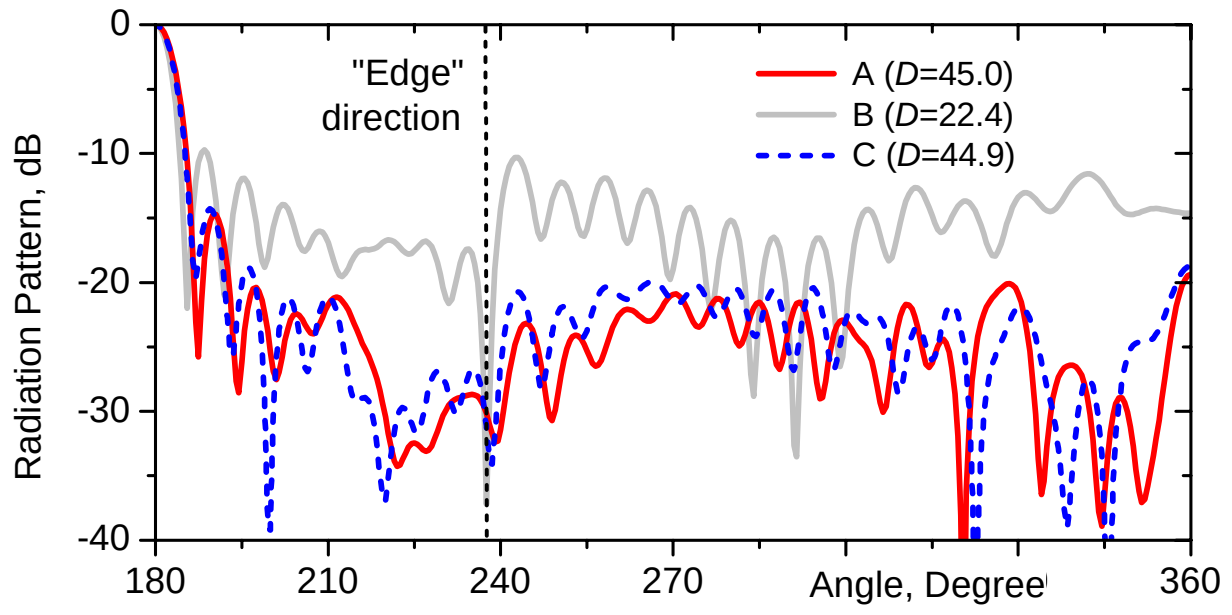

Fig. 4. Normalized far-field radiation patterns of rexolite DLAs whose parameters correspond to the relevant marks in Fig. 3(c). The directivity values for each configuration are provided in the legend.

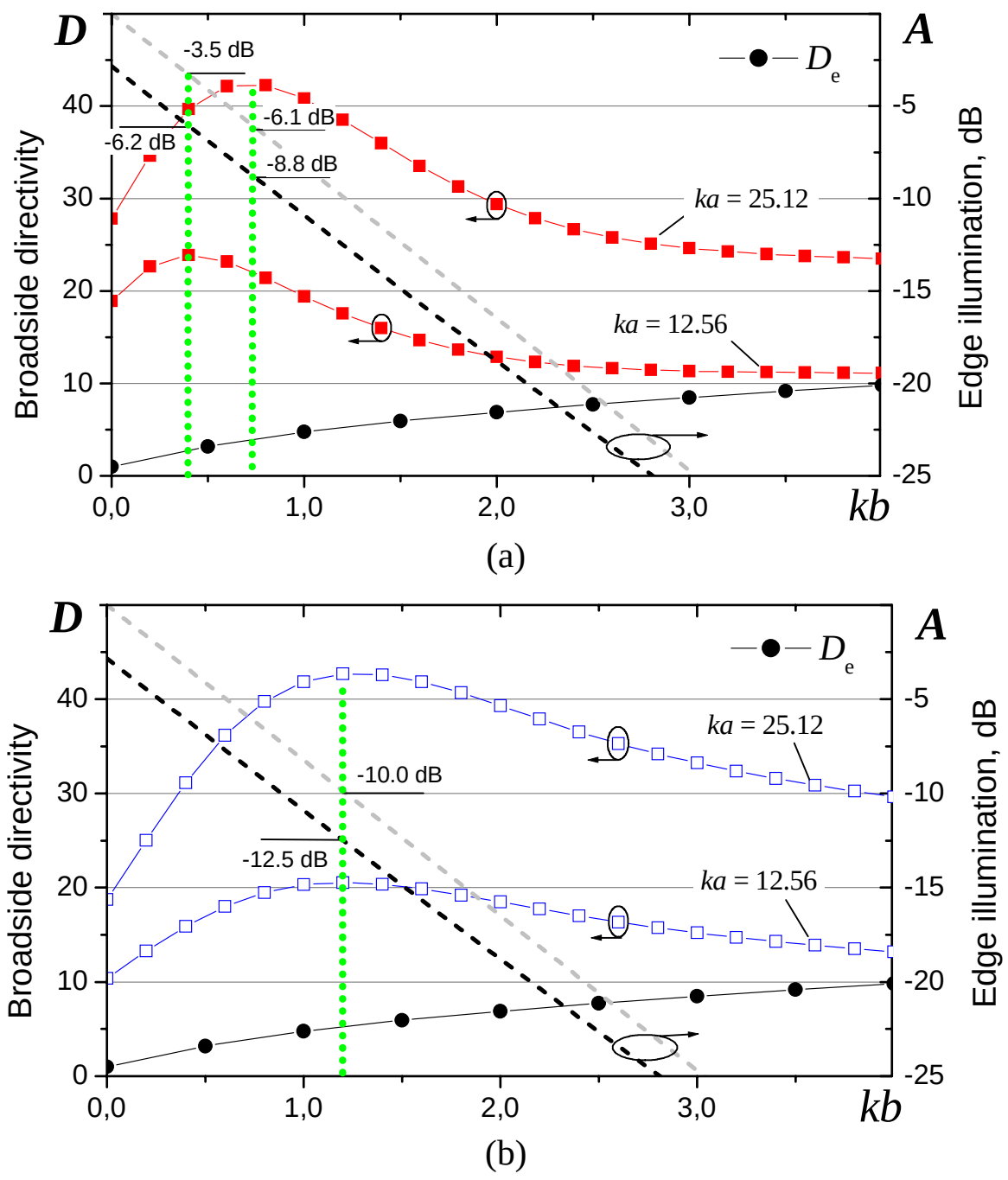

Fig. 5. The same as in Fig. 2 for the quartz DLAs ($l_1 = 0.6$, $l_2 = 1.165$): (a) $E$-polarization, (b) $H$-polarization.

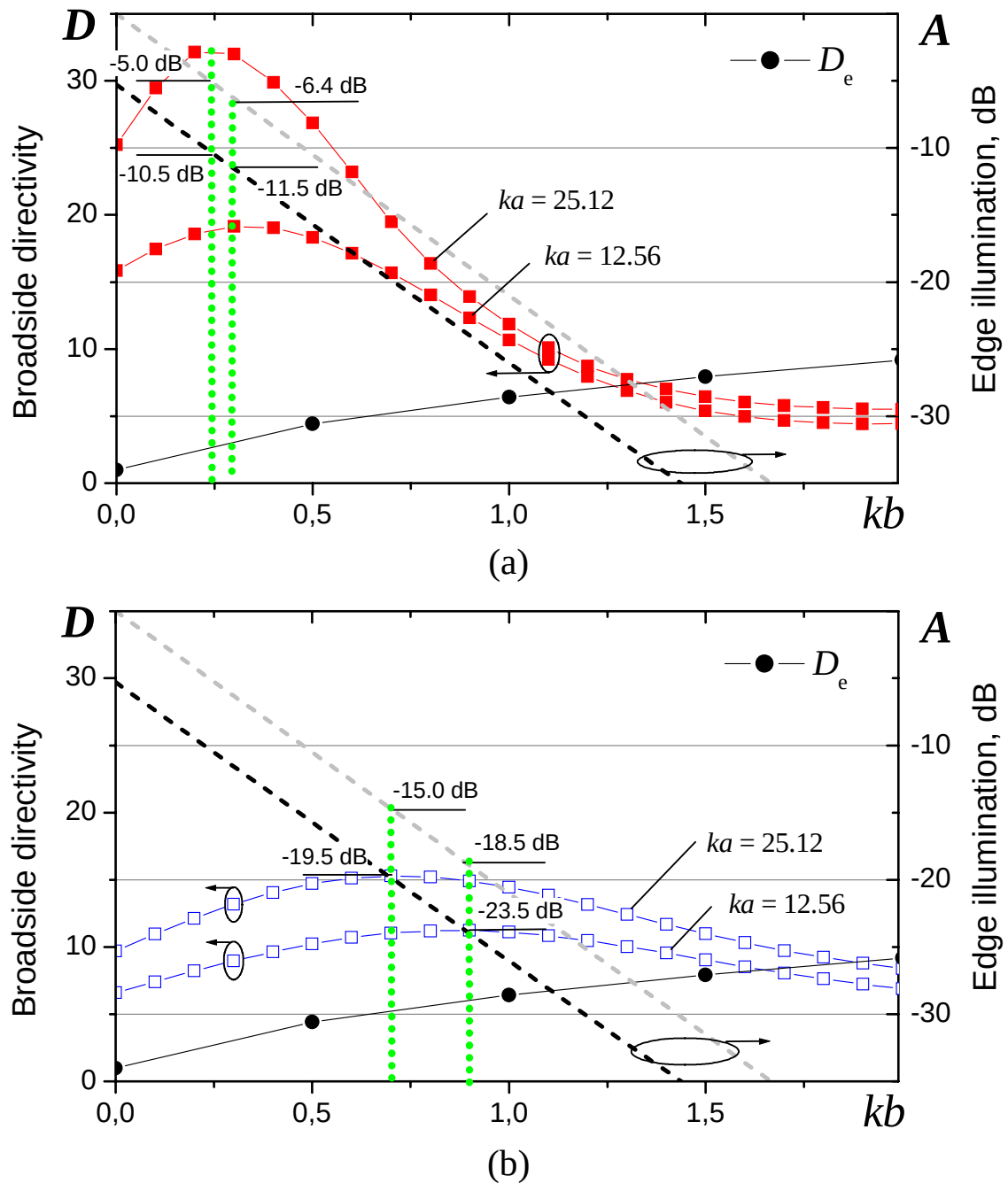

Fig. 6. The same as in Fig. 2 for the silicon DLAs ($l_1 = 0.3$, $l_2 = 1.046$): (a) *E*-polarization, (b) *H*-polarization.

This is in contrast to parabolic reflectors that operate in free space and usually have a -10 dB far-field edge taper referred as the optimal one irrespectively to antenna parameters and polarization. For a silicon lens, the above considerations are in good fit for *E*-polarization which is more sensitive for edge illumination level whereas for *H*-polarization some lower level is preferable.

The relief patterns given in Figs. 7 and 8 represent the broadside directivity versus the lens extension and the CSP aperture width for quartz and silicon lenses, respectively. As one can see, the previous conclusions about the key role of the edge illumination can be extended to denser-material lenses with one important remark. Namely, a proper choice of the edge illumination does not prevent from excitation of internal resonances that can completely spoil the directivity. This is clearly seen in Fig. 8 where deep periodic valleys running along the vertical axis are observed for a number of the lens extension values. Note that the valleys running in the top (for larger values of *kb*) are associated with Fabry-Perot effect or bouncing of the internal field that is proven by their periodicity, whereas the aperiodic ones in the figure bottom (for smaller *kb*) are associated with the so-called half-bowtie (HBT) resonances. The nature of the HBT resonances and the influence they exert upon the radiation properties of DLAs have been studied in [6, 7]. In addition to conclusions of [7], it can be noted that HBT resonances are efficiently excited only by low-directive feeds. For sharp feed beams, the influence of HBT resonances on the radiation characteristics of hemielliptic DLAs is less pronounced although still noticeable in the form of increased level of sidelobes. This is well seen in Fig. 9 presenting the radiation patterns for the silicon lenses whose configurations are marked in Fig. 8(c). Here, curve C corresponds to the resonating lens excited by the omnidirectional feed (note the splitting of the main beam and high level of sidelobes that occurs due to the HBT resonance); curve B is for the same lens excited by a directive feed (note the sidelobes that resemble the ones of the C curve); and curves A and D are for the optimal combinations of the lens and feed parameters.

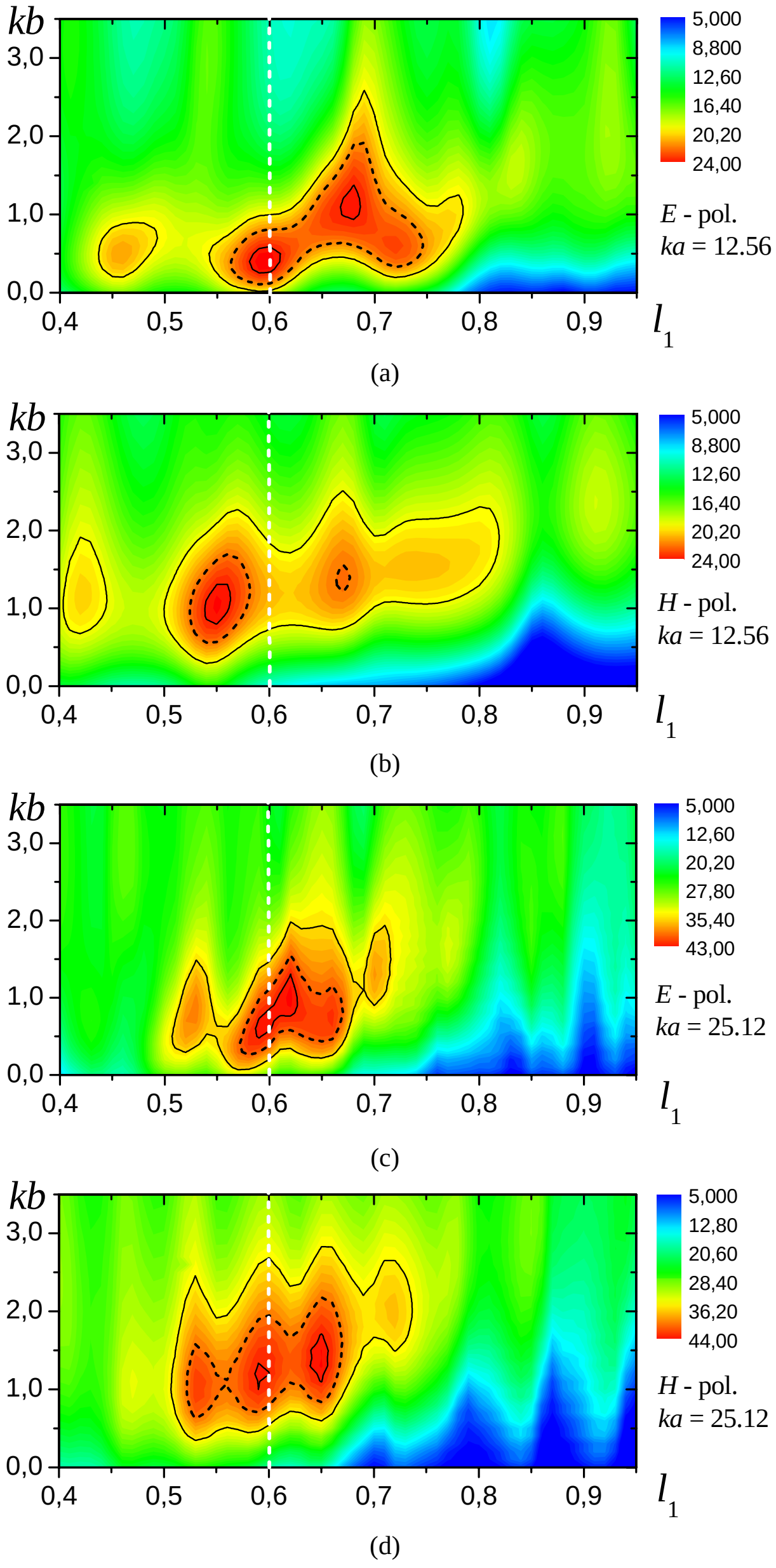

Fig. 7. The same as in Fig. 3 for the quartz DLAs.

Finally, it is worth to mention that the curves presented in Figs. 7 and 8 have been computed for the fixed frequencies and thus the values of resonant extensions will shift if the operating frequency is changed. This effect is of higher importance for systems using high-index lenses for collimating quasioptical beams, for instance in the THz time-domain spectroscopy. The wideband spectrum used in such THz systems always includes multiple frequencies that are resonant for hemielliptic lenses with any extensions and thus the HBT resonances are always involved. For antenna applications, these resonances may lead to higher-than-expected levels of side-lobes often observed in experiments, e.g. [13]. As a partial remedy against resonances, our suggestion is avoiding long extensions for silicon lenses. This is because the longer the extension the denser the spectrum of resonant values of its length and thus the more destroying the impact of the resonances on the radiation properties of DLAs.

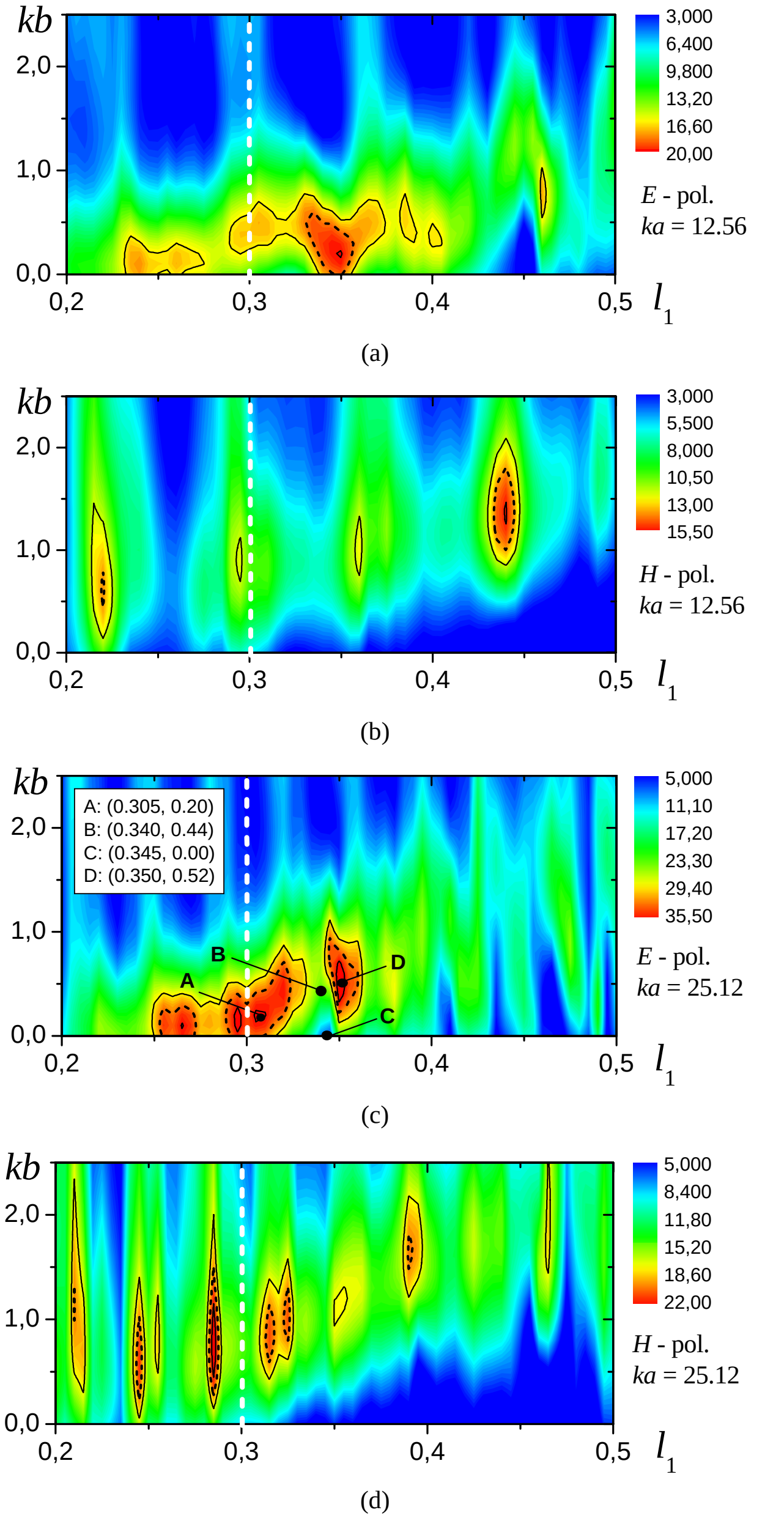

Fig. 8. The same as in Fig. 3 for the silicon DLAs. The marks A, B, C, D in (c) correspond to the far-field radiation patterns given in Fig. 9.

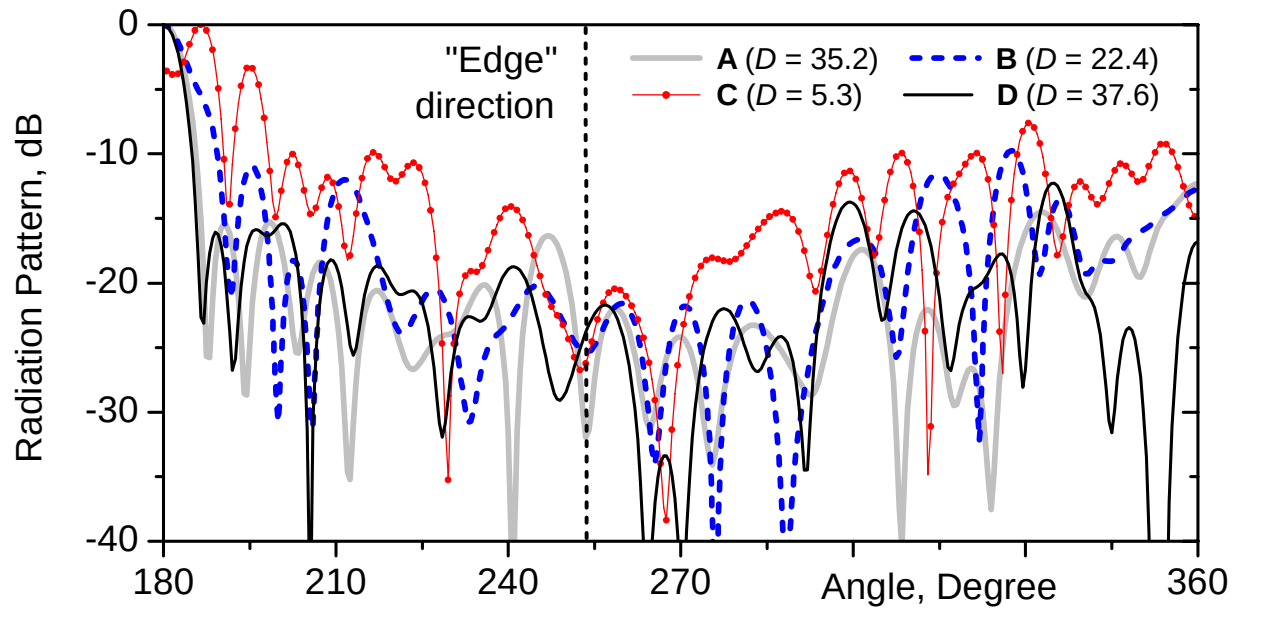

Fig. 9. Normalized far-field radiation patterns for silicon DLAs whose parameters correspond to the marks in Fig. 8(c). The values of the broadside directivity for each DLA are given in the legend.

## IV. Conclusions

The role of the edge illumination in the radiation performance of 2-D models of compact-size DLAs has been studied accurately using the in-house MBIE-based algorithm. It has been revealed that the optimal edge illumination for DLAs depends on lens material and size, as well as on the feed polarization. Moreover, it has been shown that the optimal value of -10 dB, well known for reflector antennas where far-field edge-illumination definition is common, can be still applied to DLAs if the near-field definition is used. More precisely, this recommendation is uniformly applicable to the rexolite lenses and also to the denser-material lenses in *E*-polarization. However, for *H*-polarization it must be modified in favor of -12 dB and -20 dB values for the quartz and silicon lenses, respectively. Besides, the tolerance in this value is quite large and can be estimated within the ± 5 dB range. If the far-field definition is used then the highest directivity is achieved under the edge illumination of -7÷8 dB, although the optimal value will depend on the radiation pattern of the feed used. Furthermore, this recommendation is relevant only if the frequency is far from a resonance: even a proper choice of the edge illumination does not prevent from excitation of HBT resonances that strongly affect the radiation of DLAs.